\documentclass[12pt, draftclsnofoot, onecolumn]{IEEEtran}
\usepackage{amsmath, graphics,amssymb,epsfig,subfigure}
\usepackage{cite}
\usepackage{url}
\usepackage{algorithm}
\usepackage{color,soul}
\usepackage{cases}

\def\BibTeX{{\rm B\kern-.05em{\sc i\kern-.025em b}\kern-.08em
    T\kern-.1667em\lower.7ex\hbox{E}\kern-.125emX}}

\newtheorem{theorem}{Theorem}
\newtheorem{lemma}{Lemma}

\newcommand{\be}{\begin{equation}}

\newcommand{\ee}{\end{equation}}
\newcommand{\bea}{\begin{eqnarray}}
\newcommand{\eea}{\end{eqnarray}}
\newcommand{\bdp}{\begin{displaymath}}
\newcommand{\edp}{\end{displaymath}}
\usepackage{graphicx,amsmath}

\makeatletter
\def\hlinewd#1{%
\noalign{\ifnum0=`}\fi\hrule \@height #1 \futurelet \reserved@a\@xhline}
\newcommand{\hthickline}{\hlinewd{.8pt}}

\makeatother
\begin{document}
%%%%%%%%%%%%%%%%%%%%%%%%%%%%%%%%%%%%%%%%%%%%%%%%%%%%%%%%%%%%%%%%%%%%%%%%%%%%%%%%%%%%%%%%%%%%%%% Title
\title{\huge{Resource Allocation Techniques\\for Wireless Powered Communication Networks\\with Energy Storage Constraint}}
%%%%%%%%%%%%%%%%%%%%%%%%%%%%%%%%%%%%%%%%%%%%%%%%%%%%%%%%%%%%%%%%%%%%%%%%%%%%%%%%%%%%%%%%%%%%%%
\author{\IEEEauthorblockN{\normalsize{$^\star$Hoon Lee, $^\dagger$Kyoung-Jae Lee, \textit{Member}, \textit{IEEE}, $^\star$Hanjin Kim,\\ $^*$$^\star$Bruno Clerckx, \textit{Member}, \textit{IEEE}, and $^\star$Inkyu Lee, \textit{Senior Member}, \textit{IEEE}} \\
\IEEEauthorblockA{$^\star$School of Electrical Eng., Korea University, Seoul, Korea\\
              $^\dagger$Dep. of Electronics and Control Eng., Hanbat National University, Daejeon, Korea\\
              $^*$Dep. of Electrical and Electronic Eng., Imperial College London, United Kingdom\\
    Email: \{ihun1, hanjin8612, inkyu\}@korea.ac.kr}, $^\dagger$kyoungjae@hanbat.ac.kr, $^*$b.clerckx@imperial.ac.uk \\\small}
}\maketitle \thispagestyle{empty}
%%%%%%%%%%%%%%%%%%%%%%%%%%%%%%%%%%%%%%%%%%%%%%%%%%%%%%%%%%%%%%%%%%%%%%%%%%%%%%%%%%%%%%%%%%%%%%%%%%%%%%%%Abstract
\begin{abstract}
This paper studies multi-user wireless powered communication networks, where energy constrained users charge their energy storages by scavenging energy of the radio frequency signals radiated from a hybrid access point (H-AP). The energy is then utilized for the users' uplink information transmission to the H-AP in time division multiple access mode. In this system, we aim to maximize the uplink sum rate performance by jointly optimizing energy and time resource allocation for multiple users in both infinite capacity and finite capacity energy storage cases. First, when the users are equipped with the infinite capacity energy storages, we derive the optimal downlink energy transmission policy at the H-AP. Based on this result, analytical resource allocation solutions are obtained. Next, we propose the optimal energy and time allocation algorithm for the case where each user has finite capacity energy storage. Simulation results confirm that the proposed algorithms offer about $30\ \%$ average sum rate performance gain over conventional schemes.
\end{abstract}

\section{Introduction}
Recently, radio frequency (RF) signals have been considered as a new energy source for electronic equipments \cite{PNint:12}\cite{MPinuela:13}. Unlike energy harvesting (EH) techniques based on natural energy sources such as solar or wind, the RF signal based EH systems can charge energy demanding devices whenever it is necessary. In wireless communication networks, several researches in \cite{RZhang:13,JPark:13,JPark:14,JPark:15,HLee:15} have exploited the RF signals for both wireless information transmission (WIT) and wireless energy transfer (WET), and provided simultaneous wireless information and power transfer (SWIPT) protocols in various system configurations. In the SWIPT systems, most works were confined to downlink networks, and aimed to maximize both system performance (e.g. data rate) and the harvested energy.

Wireless powered communication network (WPCN) \cite{HJu:14a,LLiu:14b,GYang:15,HJu:14,XKang:14,KHuang:14,XZhou:14b,YLing:15} is another technique which adopts the WET concept in traditional wireless communication systems. In general, the WPCN systems consist of two phases. First, in a downlink phase, devices charge their energy storages such as rechargeable batteries or supercapacitor \cite{MKaus:15} by collecting the energy of the RF signal radiated from an access point (AP). Second, in an uplink phase, the devices transmit their information signals to the AP by utilizing the energy saved in the energy storages.

In \cite{HJu:14a}, the WPCN protocol was proposed for a single antenna system where a hybrid-AP (H-AP) broadcasts the energy signal to multiple users in the downlink phase and decodes the information in the uplink phase. To facilitate multi-user detection at the H-AP, the authors in \cite{HJu:14a} employed a dynamic time division multiple access (TDMA) approach where time slots are optimally allocated to each user for maximizing the uplink throughput. By applying multiple antenna techniques \cite{Lee:06JSAC,Hakjea:09,KJLee:10,HJSung:10} to the WPCN systems, the optimal WET and WIT beamforming vectors were derived in \cite{LLiu:14b} to maximize the minimum throughput among all users. In \cite{GYang:15}, a large scale multiple antenna H-AP scenario was considered in the WPCN under an imperfect channel estimation assumption. Also, the WPCN with full duplex H-AP protocol was presented in \cite{HJu:14} and \cite{XKang:14}, where the downlink WET and the uplink WIT are performed at the same time to enhance the system performance. The authors in \cite{HJu:14} proposed joint energy and time allocation algorithms for maximizing the uplink sum rate for both perfect and imperfect self interference cancellation (SIC) scenarios. With infinite capacity energy storages at all users, \cite{HJu:14} considered a \textit{non-causal energy} system which assumes that the energy to be harvested in the future is available at the current time slot. However, this non-causal energy system may not be practical if users have finite capacity energy storages, since the energy would be insufficient for the users' uplink transmission. Furthermore, for the case of small devices such as sensor nodes which typically store the harvested energy in supercapacitors, the non-causal energy scenario is difficult to realize due to the supercapacitor's high self discharge property \cite{XKang:14}\cite{MKaus:15}. To overcome this issue, \cite{XKang:14} investigated a \textit{causal energy} system for full-duplex WPCN assuming the perfect SIC and the infinite capacity energy storage scenarios. The authors in \cite{XKang:14} optimized time allocation for the sum rate maximization and total transmission time minimization problems under uniform power allocation. For a single-user WPCN with an orthogonal frequency division multiple access technique, the optimal downlink and the uplink power allocation were obtained in \cite{XZhou:14b}.

%By separating the H-AP into a dedicated energy AP and an information AP, the optimal downlink and the uplink power allocation were obtained in \cite{XZhou:14b} for the single-user WPCN systems. Based on a stochastic geometry, the WPCN in the cellular systems were studied in \cite{KHuang:14} where energy APs are randomly deployed in traditional uplink networks, and a tradeoff between the densities of the energy APs and the information APs was characterized. The authors in \cite{YLing:15} considered a practical finite-capacity energy storage scenario in large-scale WPCN systems where the probability of successful uplink transmission was analyzed by using the stochastic geometry. Note that assuming finite-capacity energy storages at users is more practical for small devices such as mobile phones or sensor nodes.

In this paper, we study resource allocation problems in the multi-user WPCN where a H-AP broadcasts the energy RF signal to users in the downlink, and receives the users' uplink information signals by applying a dynamic TDMA approach. Unlike the conventional scheme in the non-causal energy WPCN \cite{HJu:14}, this paper considers practical causal energy systems where the users can utilize only the energy harvested at the past time slots for their uplink transmissions. In this configuration, we generalize the equal power allocation scheme in \cite{XKang:14}, and propose joint energy and time allocation methods which maximize the uplink sum rate performance in both the infinite and the finite capacity energy storage cases.

First, for the case where each user is equipped with an infinite capacity energy storage, we present an optimal downlink energy transmission policy at the H-AP. In this policy, the H-AP transfers the RF signals with the maximum power for the first few time slots, and then is turned off for the remaining time slots. Based on this result, an analytical solution for the optimal energy and time allocation is obtained. Simulation results confirm that the proposed joint optimal energy and time allocation method offers a significant performance gain compared to the equal power allocation scheme in \cite{XKang:14} which optimizes only the time durations. Also, we show that the proposed method which exploits the casual energy achieves almost identical average sum rate performance to the ideal non-causal energy system~\cite{HJu:14}.

Next, we consider a practical finite capacity energy storage case where the existing methods for the infinite energy storage case in \cite{HJu:14} and \cite{XKang:14} cannot be directly applied due to energy overflows at users. In this case, we propose an optimal resource algorithm which jointly computes the energy and time allocation. From simulation results, it is verified that the proposed optimal algorithm substantially improves the uplink sum rate performance compared to a conventional equal resource allocation scheme.

%organization
The paper is organized as follows: In Section \ref{sec:system_model}, we introduce the system model and formulate the problem for multi-user WPCN. Section \ref{sec:inf_battery} provides an analytical energy and time allocation solution for the infinite energy storage case. Also, for the finite energy storage case, the optimal resource allocation algorithm is provided in Section \ref{sec:fin_battery}. Section \ref{sec:simulation} evaluates the average sum rate performance of the proposed algorithms through numerical simulations. Finally, the paper is terminated with conclusions in Section \ref{sec:conclusion}.

\section{System Model}\label{sec:system_model}
\begin{figure}
\begin{center}
\includegraphics[width=4.0in]{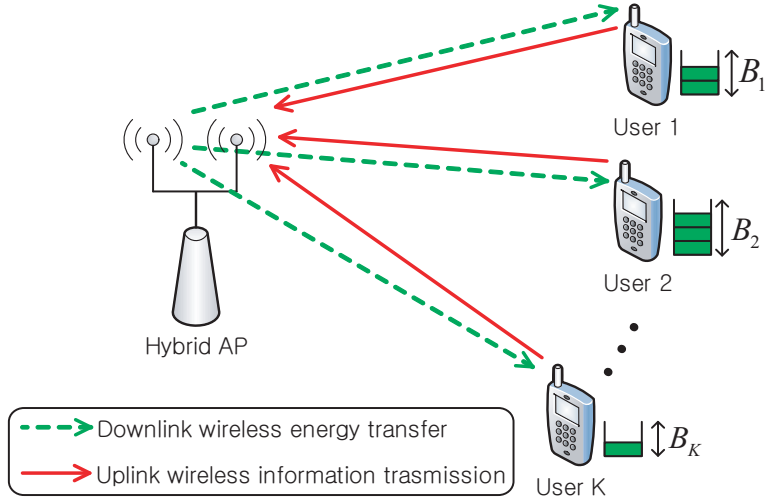}
\end{center}
%\vspace{-7mm}
\caption{Schematic diagram for multi-user WPCN}
\label{figure:system_model}
\end{figure}

We consider a $K$-user WPCN in Figure \ref{figure:system_model} where a H-AP transfers the wireless energy to single antenna users in the downlink, and at the same time, receives the users' information signals in the uplink. The H-AP has two antennas, each of which is dedicated for the downlink WET and for the uplink WIT, respectively. It is assumed that the downlink WET and the uplink WIT are scheduled over orthogonal frequency bands, i.e., the WET and the WIT signals do not interfere with each other as in \cite{XKang:14} and \cite{XZhou:14b}. In this configuration, the H-AP has a stable and fixed energy supply with average and peak power constraint $P_{A}$ and $P_{P}$, respectively,\footnote{Throughout this paper, we assume that the peak power constraint at the H-AP is larger than the average power constraint, i.e., $P_{P}>P_{A}$, without loss of generality.} while user~$i$ $(i=1,\cdots,K)$ is powered by an energy storage with capacity $B_{i}$ which is empty before transmission. To communicate with the H-AP, users first charge their energy storages by collecting the energy of the RF signal radiated from the H-AP in the downlink, and then utilize it for their uplink information transmission.

\begin{figure}
\begin{center}
\includegraphics[width=4.5in]{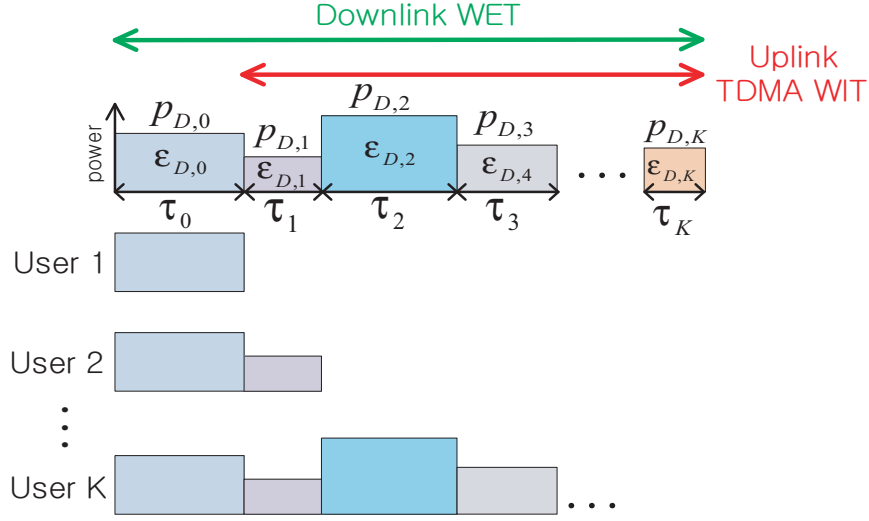}
\end{center}
%\vspace{-7mm}
\caption{Frame structure for $K$-user WPCN}
\label{figure:frame}
\end{figure}

The frame structure for the $K$-user WPCN systems is illustrated in Figure \ref{figure:frame}. For convenience, we assume that the total duration of the frame is equal to $1$ without loss of generality. The frame is divided into $K+1$ time slots. During all the $K+1$ time slots, the H-AP keeps broadcasting the RF signal to charge the users' energy storages in the downlink. For the uplink WIT, the TDMA approach is employed such that user $i$ transmits its information to the H-AP in the $i$-th time slot of duration $\tau_{i}$ for $i=1,\cdots,K$. Since the $0$-th time slot of duration $\tau_{0}$ is not scheduled to any user, it is dedicated for the downlink WET. In this setting, we consider the \textit{causal energy} scenario \cite{XKang:14} where user $i$ can only use the energy of the RF signal received at the past time slots, i.e., the time slots $0\leq j\leq i-1$, since the energy of the future RF signals is not available at the current time slot.%\footnote{The causal energy scenario is useful for sensor networks with a finite-capacity energy storage, especially with a supercapacitor which is widely used for sensor networks owing to its small form factor, fast charging cycle, and long lifetime \cite{XKang:14}.}

Assuming frequency-flat fading, the downlink and uplink channel coefficients between the H-AP and user $i$ are respectively defined as $h_{D,i}$ and $h_{U,i}$ for $i=1,\cdots,K$, which are assumed to be constant during the transmission frame. In addition, we assume that all the channel coefficients are perfectly known at the H-AP. The received signal $y_{i,j}$ at user $i$ in the $j$-th time slot with $j\neq i$ is expressed~as
\bea
    y_{i,j} = \sqrt{p_{D,j}}h_{D,i}x_{j} + n_{i,j}, \nonumber
\eea
where $p_{D,j}$ represents the downlink transmit power of the H-AP at the $j$-th time slot, $x_{j}$ stands for the energy symbol with $\mathbb{E}[|x_{j}|^{2}]=1$, and $n_{i,j}\sim\mathcal{CN}(0,\varsigma^{2}_{i,j})$ indicates the additive Gaussian noise at user $i$. %Here, the energy symbol $s_{E}$ does not carry any information, and thus we can pre-determine $s_{E}$ as an arbitrary random variable or a fixed symbol \cite{HLee:15}\cite{SLee:14}.

Then, the harvested energy from the signal $y_{i,j}$ in the $j$-th time slot is given~by
\bea
    E_{i,j} = \eta_{i}\mathbb{E}[|y_{i,j}|^{2}] = \eta_{i}g_{D,i}p_{D,j}\tau_{j}, \label{eq:E_ij}
\eea
where $\eta_{i}\in[0,1]$ denotes the energy harvesting efficiency of user $i$, and $g_{D,i}=|h_{D,i}|^{2}$ is the downlink channel gain. In (\ref{eq:E_ij}), we ignore the noise power since it is practically much smaller compared to the signal power \cite{RZhang:13}.

In the uplink information transfer, at the $i$-th time slot, only user $i$ transmits its information symbol $s_{i}\sim\mathcal{CN}(0,1)$ to the H-AP by using the energy charged in the energy storage. The received signal at the H-AP $r_{i}$ in the $i$-th time slot can be written as
%\bea
%    r_{i} = \sqrt{p_{U,i}}h_{U,i}s_{i} + \sqrt{p_{D,i}}h_{PB}s_{E} + z_{i}, \label{eq:r_i}
%\eea
\bea
    r_{i} = \sqrt{p_{U,i}}h_{U,i}s_{i} + z_{i}, \label{eq:r_i}
\eea
where $p_{U,i}$ represents the uplink transmit power of user $i$ and $z_{i}\sim\mathcal{CN}(0,\sigma_{i}^{2})$ indicates the additive Gaussian noise.

Due to the energy storage constraint at user $i$, the uplink power $p_{U,i}$ is upper bounded by
\bea
    p_{U,i} \leq \frac{B_{i}}{\tau_{i}}. \label{eq:const4}
\eea
Also, since a user can only utilize the energy harvested during the past time slots, it follows
\bea
    p_{U,i} \leq \frac{1}{\tau_{i}}\sum_{j=0}^{i-1}E_{i,j} = \frac{\eta_{i}g_{D,i}}{\tau_{i}}\sum_{j=0}^{i-1}p_{D,j}\tau_{j}. \label{eq:const5}
\eea
Let us define the downlink and uplink transmit energy as $\varepsilon_{D,i}=\tau_{i}p_{D,i}$ and $\varepsilon_{U,i}=\tau_{i}p_{U,i}$, respectively. Then, the achievable rate of user $i$ is obtained as
%It is worth noting that the energy symbol $s_{E}$ can be pre-determined, and thus it can be known to the base station in advance. Then, in order to improve each user's uplink rate, the base station cancels the interference caused by the energy symbol $s_{E}$ \cite{HLee:15}\cite{SLee:14}, i.e., $\sqrt{p_{D,i}}g_{PB}s_{E}$ in (\ref{eq:r_i}). Therefore, the achievable rate of user $i$ is obtained as
\bea
    R_{i} = \tau_{i}\log\bigg(1+\frac{g_{U,i}}{\sigma_{i}^{2}}\frac{\varepsilon_{U,i}}{\tau_{i}}\bigg), \nonumber
\eea
where $g_{U,i}=|h_{U,i}|^{2}$ is the uplink channel gain.

In this paper, we investigate the optimal energy and time allocation which maximizes the uplink sum rate. The uplink sum rate maximization problem can be formulated as
\bea
    &\max\limits_{ \{\tau_{i}\},\{\varepsilon_{D,i}\},\{\varepsilon_{U,i}\} }& \sum_{i=1}^{K} R_{i}\label{eq:P1} \\
    &\text{subject to}&\sum_{i=0}^{K} \varepsilon_{D,i} \leq P_{A}, \label{eq:const_nu} \\
    &~&\varepsilon_{D,i}\leq \tau_{i}P_{P},\ i=0,\cdots,K, \label{eq:const_new} \\
    &~&\sum_{i=0}^{K} \tau_{i} \leq 1, \label{eq:const_lambda} \\
    &~&\varepsilon_{U,i} \leq B_{i},\ i=1,\cdots,K, \label{eq:constB} \\
    &~&\varepsilon_{U,i} \leq \eta_{i}g_{D,i}\sum_{j=0}^{i-1}\varepsilon_{D,j},\ i=1,\cdots,K, \label{eq:const_beta}
\eea
where (\ref{eq:const_nu}) and (\ref{eq:const_new}) stand for the average and peak power constraint at the H-AP, respectively, and (\ref{eq:const_lambda}) denotes the total time constraint, and constraint (\ref{eq:constB}) and (\ref{eq:const_beta}) come from (\ref{eq:const4}) and (\ref{eq:const5}), respectively.

It is worth noting that the authors in \cite{XKang:14} considered a uniform downlink power allocation $p_{D,i}=\varepsilon_{D,i}/\tau_{i}=P_{A}$, $\forall i$, with the infinite capacity energy storage $B_{i}=\infty$, $\forall i$, for the casual energy WPCN systems. Also, \cite{HJu:14} studied joint energy and time allocation for the infinite capacity energy storage case in the non-casual systems, i.e., constraints (\ref{eq:constB}) and (\ref{eq:const_beta}) were not considered. Therefore, existing resource allocation methods in \cite{HJu:14} and \cite{XKang:14} cannot be directly employed to solve problem (\ref{eq:P1}). In the following sections, we provide the optimal methods to solve (\ref{eq:P1}) in two different cases. First, for the infinite capacity energy storage case, an analytical solution will be obtained. Second, we consider the practical finite capacity energy storage case, and propose an algorithm to solve (\ref{eq:P1}) optimally.

\section{Infinite Capacity Energy Storage Case}\label{sec:inf_battery}
In this section, we investigate the optimal solution of (\ref{eq:P1}) with infinite capacity energy storages at all users. By setting $B_{i}=\infty$ in problem (\ref{eq:P1}), the energy storage constraint (\ref{eq:constB}) is removed. Since the uplink rate $R_{i}$ increases with $\varepsilon_{U,i}$, the optimal uplink energy $\varepsilon_{U,i}^{\star}$ for $i=1,\cdots,K$ is given by the maximum~in~(\ref{eq:const_beta})~as
\bea
    \varepsilon_{U,i}^{\star} = \eta_{i}g_{D,i}\sum_{j=0}^{i-1}\varepsilon_{D,j}^{\star}, \nonumber
\eea
where $\varepsilon_{D,i}^{\star}$ indicates the optimal downlink energy allocation.

Substituting this into (\ref{eq:P1}), the problem can be recast to
\bea
    &\max\limits_{ \{\tau_{i}\},\{\varepsilon_{D,i}\} }& \sum_{i=1}^{K} \tau_{i}\log\bigg(1+\gamma_{i}\frac{\sum_{j=0}^{i-1}\varepsilon_{D,j}}{\tau_{i}}\bigg) \label{eq:P2} \\
    &\text{subject to}&\sum_{i=0}^{K} \varepsilon_{D,i} \leq P_{A},\ \sum_{i=0}^{K} \tau_{i} \leq 1,\nonumber \\
    &~&\varepsilon_{D,i}\leq \tau_{i}P_{P},\ i=0,\cdots,K, \nonumber
\eea
where $\gamma_{i}=\eta_{i}g_{D,i}g_{U,i}/\sigma_{i}^{2}$. Before solving problem (\ref{eq:P2}), we present the following lemma which is useful for identifying an analytical solution for (\ref{eq:P2}).
\begin{lemma} \label{lemma:lemma1}
The optimally allocated time $\{\tau_{i}^{\star}\}_{i=0}^{K}$ for problem (\ref{eq:P2}) is always greater than $0$, i.e., $\tau_{i}^{\star}>0$ for $i=0,\cdots,K$.
\end{lemma}
\begin{IEEEproof}
See Appendix \ref{appendix:appendixA}
\end{IEEEproof}
By using Lemma \ref{lemma:lemma1}, we first address the optimal downlink energy allocation policy $\{\varepsilon_{D,i}^{\star}\}_{i=0}^{K}$ in Section~\ref{sec:opteD}. Next, the computation of the optimal time allocation $\{\tau_{i}^{\star}\}_{i=0}^{K}$ will be given in Section~\ref{section:tt}.

\subsection{Optimal Downlink Energy Allocation} \label{sec:opteD}
In order to obtain $\{\varepsilon_{D,i}^{\star}\}_{i=0}^{K}$, we introduce auxiliary variables $A_{i}=\sum_{j=0}^{i}\varepsilon_{D,j}$ for $i=0,\cdots,K$ in problem (\ref{eq:P2}), which represent the transmitted energy until the $i$-th time slot. Then, problem (\ref{eq:P2}) can be rewritten as
\bea
    &\max\limits_{ \{\tau_{i}\},\{A_{i}\} }& \sum_{i=1}^{K} \log\bigg(1+\gamma_{i}\frac{A_{i-1}}{\tau_{i}}\bigg) \label{eq:P2_5} \\
    &\text{subject to}& \sum_{i=0}^{K} \tau_{i} \leq 1,\nonumber \\
    &~& A_{i}-A_{i-1}\leq \tau_{i}P_{P},\ i=0,\cdots,K, \nonumber \\
    &~& A_{i-1}\leq A_{i},\ i=0,\cdots,K+1, \label{eq:constA}
\eea
where $A_{-1}\triangleq0$, $A_{K+1}\triangleq P_{A}$, and constraint (\ref{eq:constA}) is added due to the definition of $A_{i}$.

Since the objective in (\ref{eq:P2_5}) is an increasing function of $A_{i}$, the optimal $A_{i}^{\star}$ for problem (\ref{eq:P2_5}) is determined by its maximum value as
%\bea
%    A_{i}^{\star} = \left\{ \begin{array}{l r} \label{eq:A_opt}
%                \min\{A_{i+1}^{\star},A_{i-1}^{\star}+\tau_{i}^{\star}P_{P}\}, &i=0,\cdots,K-1, \\
%                \min\{P_{A},A_{K-1}^{\star}+\tau_{K}^{\star}P_{P}\}, &i=K,~~~~~~~~~~~~~
%              \end{array}
%              \right.
%\eea
\bea
    A_{i}^{\star} =  \max\{A_{i+1}^{\star},A_{i-1}^{\star}+\tau_{i}^{\star}P_{P}\}, \label{eq:A_opt}
\eea
where $A_{-1}^{\star}\triangleq0$ and $A_{K+1}^{\star}\triangleq P_{A}$. Based on (\ref{eq:A_opt}), we provide the following theorem on the optimal downlink energy allocation solution in the infinite capacity energy storage case.
\begin{theorem} \label{theorem:theorem1}
For an arbitrarily given $L\in[0,K]$, the optimal downlink energy allocation of problem (\ref{eq:P2}) is expressed as
%\bea
%    \varepsilon_{D,i}^{\star} = \left\{ \begin{array}{l r}
%                \tau_{i}^{\star}P_{P}, &i=0,\cdots,L-1, \\
%                P_{A}-P_{P}\sum_{j=0}^{L-1}\tau_{j}^{\star}, &i=L,~~~~~~~~~~~~~ \\
%                0, &i=L+1,\cdots,K.
%              \end{array}
%              \right.
%\eea
\begin{numcases}{\varepsilon_{D,i}^{\star} =}
\tau_{i}^{\star}P_{P},~~~~~~~~~~~~~~\text{for}\ i=0,1,\cdots,L-1,\label{eq:thm1_1}\\
P_{A}-P_{P}\sum_{j=0}^{L-1}\tau_{j}^{\star},~~\text{for}\ i=L,\label{eq:thm1_2}\\
                0,~~~~~~~~~~~~~~~~~~~\text{for}\ i=L+1,\cdots,K,\label{eq:thm1_3}
\end{numcases}
where $L$ indicates the largest time slot index satisfying $\varepsilon_{D,L}^{\star}>0$. The computation of the optimal value of $L$ will be addressed later in Section \ref{section:tt}.
\end{theorem}
\begin{IEEEproof}
By subtracting $A_{i-1}^{\star}$ in both sides of (\ref{eq:A_opt}), it follows
\bea
    \varepsilon_{D,i}^{\star}=\max\{\varepsilon_{D,i}^{\star}+\varepsilon_{D,i+1}^{\star},\tau_{i}^{\star}P_{P}\}. \label{eq:proof1_1}
\eea
By using this result, we will prove (\ref{eq:thm1_1}), (\ref{eq:thm1_3}), and (\ref{eq:thm1_2}) sequentially. Consider an arbitrary time slot index $S$ such that the optimal downlink energy is positive, i.e., $\varepsilon_{D,S}^{\star}>0$. Then, we first show $\varepsilon_{D,i}^{\star}=\tau_{i}^{\star}P_{P}$ for $i=0,\cdots,S-1$ by contradiction. From (\ref{eq:proof1_1}), the optimal $\varepsilon_{D,S-1}^{\star}$ is given by either $\varepsilon_{D,S-1}^{\star}+\varepsilon_{D,S}^{\star}$ or $\tau_{S-1}^{\star}P_{P}$. If $\varepsilon_{D,S-1}^{\star}=\varepsilon_{D,S-1}^{\star}+\varepsilon_{D,S}^{\star}$, then we have $\varepsilon_{D,S}^{\star}=0$, which contradicts the fact $\varepsilon_{D,S}^{\star}>0$. Therefore, the optimal energy for the $(S-1)$-th slot is obtained as $\varepsilon_{D,S-1}^{\star}=\tau_{S-1}^{\star}P_{P}$. Utilizing this result, it is easy to prove $\varepsilon_{D,i}^{\star}=\tau_{i}^{\star}P_{P}$ for $i=0,1,\cdots,S-2$, because the optimal $\tau_{i}^{\star}$ is positive as presented in Lemma \ref{lemma:lemma1}.

Next, suppose an arbitrary time slot index $M$ $(M\leq K)$ satisfying $\varepsilon_{D,M}^{\star}=0$. Then, we now show $\varepsilon_{D,i}^{\star}=0$ for $i=M+1,M+2,\cdots,K$. From (\ref{eq:proof1_1}), it follows $\min\{\varepsilon_{D,M+1}^{\star},\tau_{M}^{\star}P_{P}\}=0$. Since this condition must be fulfilled with any positive $\tau_{M}^{\star}$, the optimal downlink energy at the $(M+1)$-th time slot is zero. Thus, it can be shown that $\varepsilon_{D,i}^{\star}=0$ for $i=M+2,M+3,\cdots,K$. Since the optimal downlink energy $\varepsilon_{D,i}^{\star}$ for $i=0,1,\cdots,L-1$ is always positive, the time slot index $M$ must be larger than $L$, i.e., $M\geq L+1$. Then, by setting $S=L$ and $M=L+1$, we verify (\ref{eq:thm1_1}) and (\ref{eq:thm1_3}).

Now, the remaining part is to prove (\ref{eq:thm1_2}). One can check that $\sum_{i=0}^{K}\varepsilon_{D,i}^{\star}=P_{A}$ is true since the objective function is non-decreasing with respect to individual $\{\varepsilon_{D,i}\}_{i=0}^{K}$. In other words, if $\sum_{i=0}^{K}\varepsilon_{D,i}<P_{A}$, then a larger uplink sum rate can be achieved by increasing some $\varepsilon_{D,i}$. Hence, to satisfy the equality $\sum_{i=0}^{K}\varepsilon_{D,i}^{\star}=P_{A}$ with (\ref{eq:thm1_1}) and (\ref{eq:thm1_3}), we have (\ref{eq:thm1_2}). Theorem \ref{theorem:theorem1} is finally proved.
\end{IEEEproof}

\begin{figure}
\centering
    \subfigure[Optimal policy]{
        \includegraphics[width=3.0in]{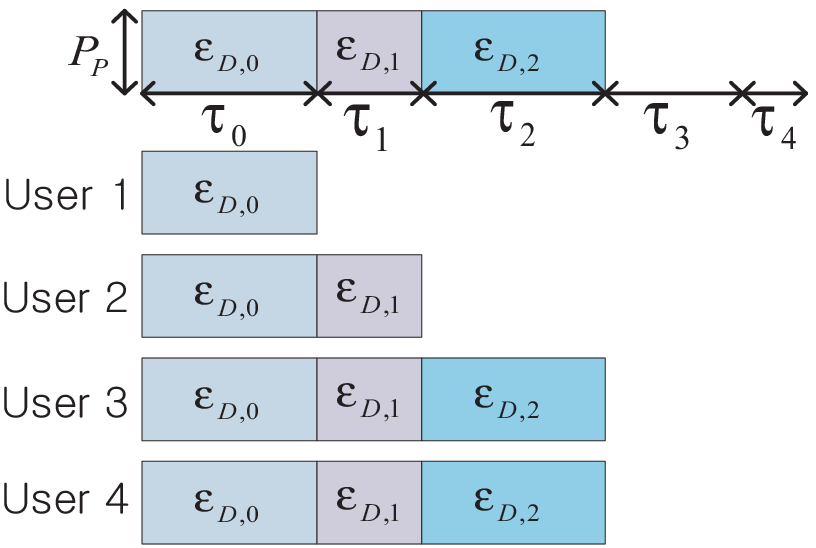}
        %\vspace{-5mm}
        %\caption{The feasible region of $\delta$ for $\bar{E}=14$}
        \label{figure:optimal}
    }
    \hspace{0mm}
    \subfigure[Suboptimal policy]{
        \includegraphics[width=3.0in]{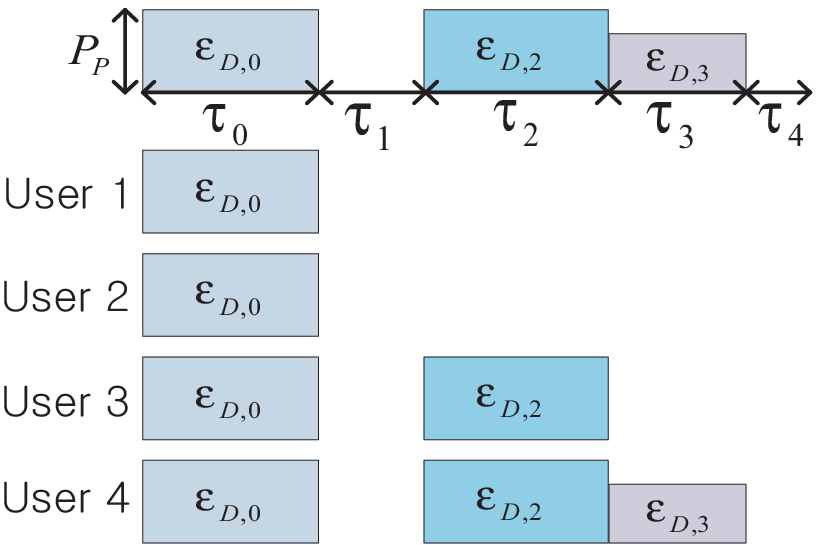}
        %\vspace{-5mm}
        %\caption{The feasible region of $\delta$ for $\bar{E}=28$}
        \label{figure:suboptimal}
    }
    \caption{Optimal and suboptimal downlink energy allocation policies with infinite capacity energy storage}
\end{figure}

Theorem \ref{theorem:theorem1} implies that for the first $i=0,1,\cdots,L-1$ time slots, the H-AP should transmit the energy RF signal with the maximum energy $\varepsilon_{D,i}^{\star}=\tau_{i}^{\star}P_{P}$, and for the $L$-th time slot, the remaining energy $P_{A}-P_{P}\sum_{j=0}^{L-1}\tau_{j}^{\star}$ is utilized. Then, the H-AP is turned off until the end of the frame. This result can be explained as follows: Due to the energy causality assumption, users can only leverage the energy harvested in the past time slots. Therefore, as shown in Figure~3~(a), it is beneficial for the H-AP to consume all available energy $P_{A}$ as soon as possible so that more energy can be transferred to users. Otherwise, the overall harvested energy of all users decreases as illustrated in Figure~3~(b), and thus the sum rate performance would be degraded. Thanks to Theorem \ref{theorem:theorem1}, we can obtain the optimal energy allocation for an arbitrarily given $L$. In the following, we proceed to determine the optimal time allocation $\{\tau_{i}^{\star}\}$ and the optimal time slot index $L^{\star}$.

\subsection{Optimal Time Allocation}\label{section:tt}
Based on Theorem \ref{theorem:theorem1}, the problem in (\ref{eq:P2}) for a given $L$ can be reformulated as
\bea
    \mathcal{R}_{L}&\triangleq&\max\limits_{ \{\tau_{i}\} } \sum_{i=1}^{L}\tau_{i}\log\left(1+\gamma_{i}P_{P}\frac{\sum_{j=0}^{i-1}\tau_{j}}{\tau_{i}}\right) + \sum_{i=L+1}^{K}\tau_{i}\log\left(1+\gamma_{i}\frac{P_{A}}{\tau_{i}}\right) \label{eq:P3} \\
    &~&\text{subject to}\ \sum_{i=0}^{K}\tau_{i}\leq 1, \nonumber
\eea
where $\mathcal{R}_{L}$ indicates the optimal value of problem (\ref{eq:P3}). To solve the problem efficiently, we introduce an auxiliary variable $T$ which splits the total time constraint in (\ref{eq:P3}) into $\sum_{i=0}^{L}\tau_{i}\leq T$ and $\sum_{i=L+1}^{K}\tau_{i} \leq 1-T$. %Note that the optimal value of $T$ will be discussed in Section \ref{sec:Topt}.

With given $T$ and $L$, the above problem can be decoupled into the following two subproblems:
\bea
    \mathcal{R}_{L}^{(1)}(T)&\triangleq&\max\limits_{ \{\tau_{i}\} } \sum_{i=1}^{L}\tau_{i}\log\left(1+\gamma_{i}P_{P}\frac{\sum_{j=0}^{i-1}\tau_{j}}{\tau_{i}}\right) \label{eq:SP1} \\
    &~&\text{subject to}\ \sum_{i=0}^{L}\tau_{i}\leq T, \nonumber
\eea
and
\bea
    \mathcal{R}_{L}^{(2)}(T)&\triangleq&\max\limits_{ \{\tau_{i}\} } \sum_{i=L+1}^{K}\tau_{i}\log\left(1+\gamma_{i}\frac{P_{A}}{\tau_{i}}\right) \label{eq:SP2} \\
    &~&\text{subject to}\ \sum_{i=L+1}^{K}\tau_{i}\leq 1-T, \nonumber
\eea
where $\mathcal{R}_{L}^{(1)}(T)$ and $\mathcal{R}_{L}^{(2)}(T)$ denote the optimal values of problems (\ref{eq:SP1}) and (\ref{eq:SP2}), respectively. Then, the optimal $T^{\star}$ and $L^{\star}$ are obtained by the maximum point of $\mathcal{R}_{L}^{(1)}(T)+\mathcal{R}_{L}^{(2)}(T)$. Hence, we should first investigate $\mathcal{R}_{L}^{(1)}(T)$ and $\mathcal{R}_{L}^{(2)}(T)$ for given $T$ and $L$, and then the optimal $T^{\star}$ and $L^{\star}$ will be determined.

\subsubsection{Optimal Solutions for (\ref{eq:SP1}) and (\ref{eq:SP2})}

We first present a solution for subproblem (\ref{eq:SP1}). One can show that the optimal solution of (\ref{eq:SP1}) is given by \cite{XKang:14}
\bea
    \tau_{i}^{\star}=
         \frac{T-\sum_{j=i+1}^{L}\tau_{j}^{\star}}{1+x_{i}},\ \text{for}\ i=0,\cdots,L.\label{eq:sol_tau1}
\eea
It is worth noting that $\tau_{i}^{\star}$ in (\ref{eq:sol_tau1}) is only affected by its future values $\{\tau_{j}^{\star}\}_{j=i+1}^{L}$, and thus it can be calculated in the reverse order. Here, $x_{i}$ for $i=0,1,\cdots,L$ is defined as
\bea
    x_{i}=\left\{ \begin{array}{l r}\label{eq:sol_x}
        0,~~~~~~~~~~~~~~~~~~~~~~\text{for}\ i=0,\\
        \frac{1}{\gamma_{i}P_{P}}\left(\frac{\gamma_{i}P_{P}-1}{w_{i}}-1\right),\ \text{for}\ i=1,\cdots,L,
    \end{array}
    \right.
\eea
where $w_{i}=\mathcal{W}\Big((\gamma_{i}P_{P}-1)\exp({-1-\sum_{j=1}^{i-1}\frac{\gamma_{j}P_{P}}{1+\gamma_{j}P_{P}x_{j}}})\Big)$ and $\mathcal{W}(\cdot)$ represents the Lambert W function \cite{RMCorless:96}. Since $x_{i}$ only depends on the previous values $\{x_{j}\}_{j=1}^{i-1}$, we can calculate $x_{1},x_{2},\cdots,x_{L}$ sequentially.

Next, we solve the second subproblem (\ref{eq:SP2}). The optimal time allocation $\{\tau_{i}^{\star}\}_{i=L+1}^{K}$ satisfies the following condition \cite{HJu:14a}:
\bea
    \frac{\gamma_{L+1}}{\tau_{L+1}^{\star}}=\frac{\gamma_{L+2}}{\tau_{L+2}^{\star}}=\cdots=\frac{\gamma_{K}}{\tau_{K}^{\star}}=C, \label{eq:C1}
\eea
where the constant $C$ is obtained as $C=\frac{\sum_{j=L+1}^{K}\gamma_{j}}{1-T}$ since the equality $\sum_{i=L+1}^{K}\tau_{i}^{\star}=1-T$ always holds \cite{HJu:14a}. Combining this and (\ref{eq:C1}), the optimal solution for (\ref{eq:SP2}) is written by
\bea
    \tau_{i}^{\star} = \frac{(1-T)\gamma_{i}}{\sum_{j=L+1}^{K}\gamma_{j}},\ \text{for}\ i=L+1,\cdots,K. \label{eq:sol_tau2}
\eea

\subsubsection{Optimal $T^{\star}$ and $L^{\star}$}\label{sec:Topt}
For solving the original problem (\ref{eq:P2}), we need to find the optimal $T^{\star}$ and $L^{\star}$, which maximize $\mathcal{R}_{L}^{(1)}(T)+\mathcal{R}_{L}^{(2)}(T)$. By substituting the optimal time allocation solutions (\ref{eq:sol_tau1}) and (\ref{eq:sol_tau2}) into the objective functions of (\ref{eq:SP1}) and (\ref{eq:SP2}), respectively, we have
\bea
    \mathcal{R}_{L}^{(1)}(T)=a_{L}T\ \text{and}\ \mathcal{R}_{L}^{(2)}(T) = (1-T)\log\Big(1+\frac{P_{A}\sum_{i=L+1}^{K}\gamma_{i}}{1-T}\Big), \nonumber
\eea
where
\bea
    a_{L}=\sum_{i=1}^{L}\frac{\prod_{j=i+1}^{L}x_{j}}{\prod_{j=i}^{L}(1+x_{j})}\log(1+\gamma_{i}P_{P}x_{i}).\nonumber
\eea
It is easy to verify that $\mathcal{R}_{L}^{(1)}(T)+\mathcal{R}_{L}^{(2)}(T)$ is a concave function with respect to $T$, and thus the optimal $T^{\star}$ can be determined from the stationary point $\tilde{T}$, which is computed~as
\bea
    \tilde{T}=\frac{P_{A}\sum_{i=L+1}^{K}\gamma_{i}}{1/\mathcal{W}(-\exp({-1-a_{L}}))+1}+1. \nonumber
\eea

Now, we check the feasible region of $T$ with a given $L$. Due to the fact $T=\sum_{i=0}^{L}\tau_{i}^{\star}$, we can rewrite $\varepsilon_{D,L}^{\star}$ in (\ref{eq:thm1_2}) as $  \varepsilon_{D,L}^{\star}=P_{A}-P_{P}(T-\tau_{L}^{\star})$. Since $\varepsilon_{D,L}^{\star}$ is positive, it follows
\bea
    T\leq \frac{P_{A}}{P_{P}} + \tau_{L}^{\star} = \frac{P_{A}}{P_{P}}\Big(1+\frac{1}{x_{L}}\Big). \nonumber
\eea
Also, from the peak power constraint $\varepsilon_{D,L}^{\star}\leq\tau_{L}^{\star}P_{P}$, $T$ should be lower bounded by $T\geq P_{A}/P_{P}$. Therefore, a closed-form expression for $T^{\star}$ is calculated by
\bea
    T^{\star} = \left\{ \begin{array}{l r} \label{eq:T_opt}
        \frac{P_{A}}{P_{P}},&\text{if}\ \tilde{T}<\frac{P_{A}}{P_{P}}, ~~~~~~~~~~~ \\
        \frac{P_{A}}{P_{P}}\Big(1+\frac{1}{x_{L}}\Big),&\text{if}\ \tilde{T}>\frac{P_{A}}{P_{P}}\Big(1+\frac{1}{x_{L}}\Big), \\
        \tilde{T},&\text{otherwise},~~~~~~~~~~~~
    \end{array}
    \right.\nonumber\\
    =\min\bigg\{\max\Big\{\tilde{T},\frac{P_{A}}{P_{P}}\Big\},\frac{P_{A}}{P_{P}}\Big(1+\frac{1}{x_{L}}\Big)\bigg\}.
\eea

For a given $L$, we can attain the optimal value $\mathcal{R}_{L}$ of problem (\ref{eq:P3}) as $\mathcal{R}_{L}=\mathcal{R}_{L}^{(1)}(T^{\star})+\mathcal{R}_{L}^{(2)}(T^{\star})$. Then, the optimal time slot index $L^{\star}$ is determined as
\bea
    L^{\star}=\arg\max_{0\leq L\leq K}\mathcal{R}_{L}. \label{eq:opt_L}
\eea
Note that in order to compute $\mathcal{R}_{l}$ for $l=0,\cdots,K$, we only need $\{x_{i}\}_{i=1}^{K}$ in (\ref{eq:sol_x}), which can be calculated in advance. After we obtain $L^{\star}$ and the corresponding $T^{\star}$, the optimal resource allocation solution $\{\tau_{i}^{\star},\varepsilon_{D,i}^{\star}\}_{i=0}^{K}$ can be obtained from (\ref{eq:thm1_1})-(\ref{eq:thm1_3}), (\ref{eq:sol_tau1}) and (\ref{eq:sol_tau2}). An algorithm for solving the uplink sum rate maximization problem in (\ref{eq:P2}) is summarized below.
 \renewcommand{\arraystretch}{1.}
 \begin{center}
 \begin{tabular}{l}
 \hthickline
 Algorithm 1: Optimal algorithm with infinite capacity energy storage\\
 \hthickline
 Compute $\{x_{i}\}_{i=1}^{K}$ from (\ref{eq:sol_x}).\\
 Obtain $L^{\star}$ from (\ref{eq:opt_L}) and the corresponding $T^{\star}$ from (\ref{eq:T_opt}).\\
 Compute $\{\tau_{i}^{\star}\}_{i=0}^{K}$ from (\ref{eq:sol_tau1}) and (\ref{eq:sol_tau2}) with $T=T^{\star}$ and $L=L^{\star}$.\\
 Compute $\{\varepsilon_{D,i}^{\star}\}_{i=0}^{K}$ from (\ref{eq:thm1_1})-(\ref{eq:thm1_3}) with $T=T^{\star}$ and $L=L^{\star}$.\\
 \hthickline
 \end{tabular}
 \end{center}
%\begin{table}
%\centering \caption{Algorithm 1: Optimal algorithm with infinite-capacity energy storage} \label{table:table1}
%  \begin{tabular}{ | l |}
%    \hline
%    Compute $\{x_{i}\}_{i=1}^{K}$ from (\ref{eq:sol_x}).\\
%    Obtain $L^{\star}$ from (\ref{eq:opt_L}) and the corresponding $T^{\star}$ from (\ref{eq:T_opt}).\\
%    Compute $\{\tau_{i}^{\star}\}_{i=0}^{K}$ from (\ref{eq:sol_tau1}) and (\ref{eq:sol_tau2}) with $T=T^{\star}$ and $L=L^{\star}$.\\
%    Compute $\{\varepsilon_{D,i}^{\star}\}_{i=0}^{K}$ from (\ref{eq:thm1_1})-(\ref{eq:thm1_3}) with $T=T^{\star}$ and $L=L^{\star}$.\\
%    \hline
%  \end{tabular}
%\end{table}
%\begin{algorithm}
%Optimal algorithm with infinite capacity energy storage\\
%Compute $\{x_{i}\}_{i=1}^{K}$ from (\ref{eq:sol_x}).\\
%Obtain $L^{\star}$ from (\ref{eq:opt_L}) and the corresponding $T^{\star}$ from (\ref{eq:T_opt}).\\
%Compute $\{\tau_{i}^{\star}\}_{i=0}^{K}$ from (\ref{eq:sol_tau1}) and (\ref{eq:sol_tau2}) with $T=T^{\star}$ and $L=L^{\star}$.\\
%Compute $\{\varepsilon_{D,i}^{\star}\}_{i=0}^{K}$ from (\ref{eq:thm1_1})-(\ref{eq:thm1_3}) with $T=T^{\star}$ and $L=L^{\star}$.
%\end{algorithm}

\section{Finite Capacity Energy Storage Case}\label{sec:fin_battery}
In this section, we propose the optimal energy and time allocation algorithm for the practical finite energy storage capacity scenario, i.e., $B_{i}<\infty,\forall i$, by investigating the original problem in (\ref{eq:P1}). It is worth noting that problem (\ref{eq:P1}) is convex and satisfies the Slater's condition, and thus the duality gap is zero. As a result, we can apply the Lagrange duality method to solve (\ref{eq:P1}) optimally. The Lagrangian of (\ref{eq:P1}) is expressed as
\bea
    \mathcal{J} = \sum_{i=1}^{K}\tau_{i}\log\bigg(1+\frac{g_{U,i}}{\sigma_{i}^{2}}\frac{\varepsilon_{U,i}}{\tau_{i}}\bigg) +\nu\bigg(P_{A}-\sum_{i=0}^{K}\varepsilon_{D,i}\bigg)+ \lambda\bigg(1-\sum_{i=0}^{K}\tau_{i}\bigg)\nonumber\\
     + \sum_{i=1}^{K}\beta_{i}\bigg(\eta_{i}g_{D,i}\sum_{j=0}^{i-1}\varepsilon_{D,j}-\varepsilon_{U,i}\bigg),~~~~~~~~~~~~~~~~~~~~~~~~~~~~~~~~~~~~\nonumber
\eea
where $\nu$, $\lambda$, and $\{\beta_{i}\}_{i=1}^{K}$ represent the dual variables corresponding to the constraint (\ref{eq:const_nu}), (\ref{eq:const_lambda}), and (\ref{eq:const_beta}), respectively.

Then, the dual function $\mathcal{G}(\nu,\lambda,\{\beta_{i}\}_{i=1}^{K})$ is given by
\bea
    \mathcal{G}(\nu,\lambda,\{\beta_{i}\}_{i=1}^{K}) = \max\limits_{\{\tau_{i}\},\{\varepsilon_{D,i}\},\{\varepsilon_{U,i}\}} \mathcal{J}~~~~~~~~~~~~~~~~\label{eq:dual_func} \\
    \text{subject to}\ 0\leq \varepsilon_{D,i} \leq \tau_{i}P_{P},\ i=0,\cdots,K, \nonumber \\
    0\leq \varepsilon_{U,i} \leq B_{i},\ i=1,\cdots,K.~~~\nonumber
\eea
Therefore, to solve the dual problem of (\ref{eq:P1}), which is defined as $\min_{\nu,\lambda,\{\beta_{i}\}}\mathcal{G}(\nu,\lambda,\{\beta_{i}\}_{i=1}^{K})$, we first consider the maximization problem in (\ref{eq:dual_func}) with the given dual variables. Then, the optimal dual solutions $\nu^{\star}$, $\lambda^{\star}$, and $\{\beta_{i}^{\star}\}_{i=1}^{K}$ can be obtained by solving the dual problem.

It is worthwhile to note that the Lagrangian $\mathcal{J}$ can be rewritten by $\mathcal{J}=\sum_{i=0}^{K}\mathcal{J}_{i}+\lambda+\nu P_{A}$, where
\bea
    \mathcal{J}_{i}=\left\{ \begin{array}{l r}
        -\lambda\tau_{0}+\bigg(\sum_{i=1}^{K}\eta_{i}g_{D,i}\beta_{i}-\nu\bigg)\varepsilon_{D,0},&\text{for}\ i=0,~~~~~~~~~\nonumber \\
        \tau_{i}\log\bigg(1+\frac{g_{U,i}}{\sigma_{i}^{2}}\frac{\varepsilon_{U,i}}{\tau_{i}}\bigg)-\lambda\tau_{i}+\bigg(\sum_{j=i+1}^{K}\eta_{j}g_{D,j}\beta_{j}-\nu\bigg)\varepsilon_{D,i}-\beta_{i}\varepsilon_{U,i},&\text{for}\ i=1,\cdots,K.
    \end{array}
    \right.
\eea
Since $\mathcal{J}_{i}$ depends only on $\tau_{i}$, $\varepsilon_{D,i}$, and $\varepsilon_{U,i}$, problem (\ref{eq:dual_func}) can be decomposed into $K+1$ independent optimization problems. The $i$-th problem for $i=0,\cdots,K$ is given by
\bea
    \max\limits_{\tau_{i},\varepsilon_{D,i},\varepsilon_{U,i}}\mathcal{J}_{i}~~~~~~~~~~~~~~~~~~\label{eq:dual_decomp} \\
    \text{subject to}\ \ 0\leq \varepsilon_{D,i} \leq \tau_{i}P_{P},\nonumber \\
                       0\leq \varepsilon_{U,i} \leq B_{i}.~~~\nonumber
\eea
In the following lemma, we provide solutions of problem (\ref{eq:dual_decomp}) $\{\tilde{\tau_{i}}\}_{i=0}^{K}$, $\{\tilde{\varepsilon}_{D,i}\}_{i=0}^{K}$, and $\{\tilde{\varepsilon}_{U,i}\}_{i=1}^{K}$ which maximize the Lagrangians $\mathcal{J}_{i}$ for $i=0,\cdots,K$.

\begin{lemma} \label{lemma:lemma2}
With given $\nu$, $\lambda$, and $\{\beta_{i}\}_{i=1}^{K}$, the solutions $\{\tilde{\tau}_{i}\}_{i=0}^{K}$, $\{\tilde{\varepsilon}_{D,i}\}_{i=0}^{K}$, and $\{\tilde{\varepsilon}_{U,i}\}_{i=1}^{K}$ which maximize the Lagrangian is expressed by
\bea
    \tilde{\tau}_{i}\!\!&=&\!\!\left\{ \begin{array}{l r}
        1, &\text{for}\ i=0\ \text{and}\ P_{P}\zeta_{0}-\lambda>0, \\
        0, &\text{for}\ i=0\ \text{and}\ P_{P}\zeta_{0}-\lambda\leq0, \\
        \dfrac{-g_{U,i}/\sigma_{i}^{2}}{1/b_{i}+1}\tilde{\varepsilon}_{U,i}, &\text{for}\ i=1\cdots,K,\!~~~~~~~~~~~~~~
    \end{array}
    \right. \label{eq:tilde_tau} \\
    \tilde{\varepsilon}_{D,i}\!\!&=&\!\!\left\{ \begin{array}{l r}
        \tilde{\tau}_{i}P_{P}, &\text{if}\ \zeta_{i} > 0,~ \\
        0,&\text{otherwise}~,
    \end{array}
    \right.\label{eq:tilde_eD} \\
    \tilde{\varepsilon}_{U,i}\!\!&=&\!\!\left\{ \begin{array}{l r}
        B_{i},&\text{if}\ \beta_{i}=0\ \text{or}\ \beta_{i}<-g_{U,i}b_{i}/\sigma_{i}^{2},~~~~~~ \\
        0,&\text{if}\ \beta_{i}=g_{U,i}/\sigma_{i}^{2}\ \text{or}\ \beta_{i}>-g_{U,i}b_{i}/\sigma_{i}^{2}, \\
        z_{i},&\text{otherwise},~~~~~~~~~~~~~~~~~~~~~~~~~~~~~%\beta_{i}=-g_{U,i}b_{i}/\sigma_{i}^{2},~~~~~~~~~~~~~~
    \end{array}
    \right.\label{eq:tilde_eU}
    %\tilde{\varepsilon}_{U,i}\!\!&=&\!\!\min\bigg\{\bigg(\frac{1}{\beta_{i}}-\frac{1}{h_{U,i}}\bigg)\tilde{\tau}_{i},B_{i}\bigg\},\ i=1,\cdots,K, \nonumber
\eea
where $b_{i}\triangleq\mathcal{W}(-\exp({-1-\lambda+P_{P}\zeta_{i}}))$ for $i=1,\cdots,K$, $\zeta_{i}\triangleq(\sum_{j=i+1}^{K}\eta_{j}g_{D,j}\beta_{j}-\nu)^{+}$ for $i=0,\cdots,K$ with $(x)^{+}\triangleq\max\{0,x\}$, and $z_{i}\leq B_{i}$ for $i=1,\cdots,K$ is a non-negative number which will be determined later. Also, to guarantee feasible $\tilde{\tau}_{i}$ and $\tilde{\varepsilon}_{U,i}$, we should satisfy the dual constraint $P_{P}\zeta_{i}-\lambda \leq 0$ for $i=1,\cdots,K-1$ and $\beta_{i}\leq g_{U,i}/\sigma_{i}^{2}$ for $i=1,\cdots,K$.
\end{lemma}
\begin{IEEEproof}
Please see Appendix \ref{appendix:appendixB}.
\end{IEEEproof}

From Lemma \ref{lemma:lemma2}, we can obtain the primal optimal solutions maximizing $\mathcal{J}$ with the given dual variables $\nu$, $\lambda$, and $\{\beta_{i}\}_{i=1}^{K}$. Also, Lemma \ref{lemma:lemma2} implies that the optimal downlink energy allocation policy in (\ref{eq:tilde_eD}) is similar to that of the infinite capacity energy storage case, i.e., the H-AP transmits the energy RF signal with its maximum energy $\tilde{\varepsilon}_{D,i}=\tilde{\tau}_{i}P_{P}$ during the first few time slots, and then it will be turned off. Based on Lemma \ref{lemma:lemma2}, the dual problem can be written~by
\bea
    \min\limits_{\nu,\lambda,\{\beta_{i}\}} \mathcal{G}(\nu,\lambda,\{\beta_{i}\}_{i=1}^{K})~~~~~~~~~~~~~~~~~~~~~~~~\nonumber \\
    \text{subject to}\ \nu\geq 0,\ \lambda\geq 0,~~~~~~~~~~~~~~~~~~~~~~\nonumber \\
    0\leq\beta_{i}\leq \frac{g_{U,i}}{\sigma_{i}^{2}},\ i=1,\cdots,K,~~~~~\label{eq:dual_const1} \\
    P_{P}\zeta_{i}-\lambda \leq 0,\ i=1,\cdots,K-1,\label{eq:dual_const2}
\eea
where constraint (\ref{eq:dual_const1}) and (\ref{eq:dual_const2}) come from Lemma \ref{lemma:lemma2}.

The optimal dual solutions $\nu^{\star}$, $\lambda^{\star}$, and $\{\beta_{i}^{\star}\}_{i=1}^{K}$ can be efficiently determined by subgradient methods, e.g., the ellipsoid method \cite{Boyd:04}. Note that the subgradient of the dual function $\mathcal{G}(\nu,\lambda,\{\beta_{i}\}_{i=1}^{K})$ is computed as $\boldsymbol{\mu}=[\mu_{\nu},\mu_{\lambda},\mu_{\beta_{1}},\cdots,\mu_{\beta_{K}}]$, where $\mu_{\nu}=P_{A}-\sum_{i=0}^{K}\tilde{\varepsilon}_{D,i}$, $\mu_{\lambda}=1-\sum_{i=0}^{K}\tilde{\tau}_{i}$, and $\mu_{\beta_{i}}=\eta_{i}g_{D,i}\sum_{j=0}^{i-1}\tilde{\varepsilon}_{D,j}-\tilde{\varepsilon}_{U,i}$ for $i=1,\cdots,K$. In addition, we need the subgradient of the constraint in (\ref{eq:dual_const2}), which is not easy to derive due to the definition of $\zeta_{i}$. To this end, we introduce the following lemma which provides the equivalent condition of (\ref{eq:dual_const2}).
\begin{lemma}
The constraint in (\ref{eq:dual_const2}) for $i=1,\cdots,K-1$ is equivalent to
\bea
    P_{P}\bigg(\sum_{j=2}^{K}\eta_{j}g_{D,j}\beta_{j}-\nu\bigg)-\lambda\leq 0. \label{eq:cond1}
\eea
\end{lemma}
\begin{IEEEproof}
We prove this lemma for two cases $\zeta_{1}=0$ and $\zeta_{1}=\sum_{j=2}^{K}\eta_{j}g_{D,j}\beta_{j}-\nu>0$.
First, if $\zeta_{1}=0$, i.e., $\sum_{j=2}^{K}\eta_{j}g_{D,j}\beta_{j}-\nu\leq 0$, then we have $\zeta_{i}=0$ for $i=2,\cdots,K-1$, since $\beta_{i}$ is a non-negative number. In this case, both the constraint (\ref{eq:dual_const2}) and the condition (\ref{eq:cond1}) become equivalent to $\lambda\geq 0$, i.e., (\ref{eq:dual_const2}) and (\ref{eq:cond1}) are the same. On the other hand, for the second case of $\zeta_{1}=\sum_{j=2}^{K}\eta_{j}g_{D,j}\beta_{j}-\nu>0$, the condition (\ref{eq:cond1}) is equivalent to (\ref{eq:dual_const2}) for $i=1$. Also, if the condition (\ref{eq:cond1}) is satisfied, the inequalities (\ref{eq:dual_const2}) for $i=2,\cdots,K-1$ are directly obtained since $\beta_{i}\geq 0$. This completes the proof.
\end{IEEEproof}

After we compute $\nu^{\star}$, $\lambda^{\star}$, and $\{\beta_{i}^{\star}\}_{i=1}^{K}$, it still remains to find $z_{i}$ in (\ref{eq:tilde_eU}). This can be determined by the complementary slackness condition of problem (\ref{eq:P1}) as
\bea
    \beta_{i}^{\star}\bigg(\eta_{i}g_{D,i}\sum_{j=0}^{i-1}\varepsilon_{D,j}^{\star}-\varepsilon_{U,i}^{\star}\bigg)=0, \label{eq:comslack}
\eea
where $\{\varepsilon_{D,i}^{\star}\}_{i=0}^{K}$ and $\{\varepsilon_{U,i}^{\star}\}_{i=1}^{K}$ indicate the optimal solution for problem (\ref{eq:P1}) with the optimal dual variables $\nu^{\star}$, $\lambda^{\star}$, and $\{\beta_{i}^{\star}\}_{i=1}^{K}$. Let us define the set $\mathcal{S}$ as $\mathcal{S}=\{i|\beta_{i}^{\star}>0\}$. Then, to satisfy (\ref{eq:comslack}), the optimal uplink energy allocation is given by $\varepsilon_{U,i}^{\star} = \eta_{i}g_{D,i}\sum_{j=0}^{i-1}\varepsilon_{D,j}^{\star}$ for $i\in\mathcal{S}$. It is worth noting that from (\ref{eq:tilde_eU}), we have $\varepsilon_{U,i}^{\star}=B_{i}$ for $i\in\mathcal{S}^{c}$, where $\mathcal{S}^{c}$ is the complementary set of $\mathcal{S}$.

Combining these results and (\ref{eq:tilde_tau}), the optimal time allocation solution $\{\tau_{i}^{\star}\}_{i=1}^{K}$ can be written by $\tau_{i}^{\star}=\dfrac{-g_{U,i}/\sigma_{i}^{2}}{1/b_{i}^{\star}+1}\varepsilon_{U,i}^{\star}$ for $i=1,\cdots,K$, where $b_{i}^{\star}$ is equal to $b_{i}$ with the optimal dual variables. Also, one can prove that $\tau_{0}^{\star}=\varepsilon_{D,0}^{\star}/P_{P}$, since otherwise we have $\varepsilon_{D,0}^{\star}=0$ from (\ref{eq:tilde_eD}) which implies $\varepsilon_{D,i}^{\star}=0$ for $i=1,\cdots,K$ due to the fact $\beta_{i}^{\star}\geq 0$, and obviously, this is not the optimal solution. With $\tau_{i}=\tau_{i}^{\star}$ and $\varepsilon_{U,i}=\varepsilon_{U,i}^{\star},\forall i$, problem (\ref{eq:P1}) becomes a linear programming (LP) in terms of $\{\varepsilon_{D,i}\}_{i=0}^{K}$. The optimal downlink energy allocation $\{\varepsilon_{D,i}^{\star}\}_{i=0}^{K}$ can be efficiently identified by solving this LP via the simplex algorithm or the interior-point method \cite{Boyd:04}. We summarize the overall algorithm for the finite capacity energy storage case below.\footnote{The sum rate maximization problem for the infinite storage case can also be solved via Algorithm 2. However, Algorithm 1 is still meaningful because of the computational complexity. It is worth noting that Algorithms 1 and 2 require $\mathcal{O}(K)$ and $\mathcal{O}(K^{3})$ computations, respectively \cite{Boyd:04}. Therefore, in the special case of $B_{i}=\infty$, Algorithm~1 is more efficient than Algorithm~2.}
 \renewcommand{\arraystretch}{1.}
 \begin{center}
 \begin{tabular}{l}
 \hthickline
 Algorithm 2: Optimal algorithm with finite capacity energy storage\\
 \hthickline
 Initialize $\nu$, $\lambda$, and $\{\beta_{i}\}_{i=1}^{K}$.\\
 Repeat\\
 ~~~~Compute $\{\tilde{\tau}\}_{i=0}^{K}$, $\{\tilde{\varepsilon}_{D,i}\}_{i=0}^{K}$, and $\{\tilde{\varepsilon}_{U,i}\}_{i=1}^{K}$ from (\ref{eq:tilde_tau})-(\ref{eq:tilde_eU}).\\
 ~~~~Update $\nu$, $\lambda$, and $\{\beta_{i}\}_{i=1}^{K}$ by using the ellipsoid method.\\
 Until convergence\\
 Obtain $\{\varepsilon_{D,i}^{\star}\}_{i=0}^{K}$ by solving the problem (\ref{eq:P1}) with $\tau_{i}=\tau_{i}^{\star}$ $(i=0,\cdots,K)$\\and $\varepsilon_{U,i}=\varepsilon_{U,i}^{\star}$ $(i=1,\cdots,K)$.\\
 \hthickline
 \end{tabular}
 \end{center}
% \begin{table}
%\centering \caption{ Algorithm 2: Optimal algorithm with finite-capacity energy storage} \label{table:table2}
%  \begin{tabular}{ | l |}
%    \hline
% Initialize $\nu$, $\lambda$, and $\{\beta_{i}\}_{i=1}^{K}$.\\
% Repeat\\
% ~~~~Compute $\{\tilde{\tau}\}_{i=0}^{K}$, $\{\tilde{\varepsilon}_{D,i}\}_{i=0}^{K}$, and $\{\tilde{\varepsilon}_{U,i}\}_{i=1}^{K}$ from (\ref{eq:tilde_tau})-(\ref{eq:tilde_eU}).\\
% ~~~~Update $\nu$, $\lambda$, and $\{\beta_{i}\}_{i=1}^{K}$ by using the ellipsoid method.\\
% Until convergence\\
% Obtain $\{\varepsilon_{D,i}^{\star}\}_{i=0}^{K}$ by solving problem (\ref{eq:P1}) with $\tau_{i}=\tau_{i}^{\star}$ $(i=0,\cdots,K)$ and $\varepsilon_{U,i}=\varepsilon_{U,i}^{\star}$ $(i=1,\cdots,K)$.\\
%    \hline
%  \end{tabular}
%\end{table}

\section{Simulation Results}\label{sec:simulation}
In this section, we provide numerical results evaluating the average sum rate performance of the proposed algorithms for the infinite capacity and finite capacity energy storage cases. In the simulations, we set the energy harvesting efficiency $\eta_{i}$ as $\eta_{i}=0.7$, $\forall i$, and the noise power at the H-AP $\sigma_{i}^{2}$ as $\sigma_{i}^{2}=-50\ \text{dBm}$, $\forall i$. Also, it is assumed that all users' energy storages have the same capacity, i.e., $B_{i}=B,\forall i$, and we employ the Rayleigh fading channel model with $30\ \text{dB}$ average signal attenuation from the H-AP to all users.

\begin{figure}
\begin{center}
\includegraphics[width=4.5in]{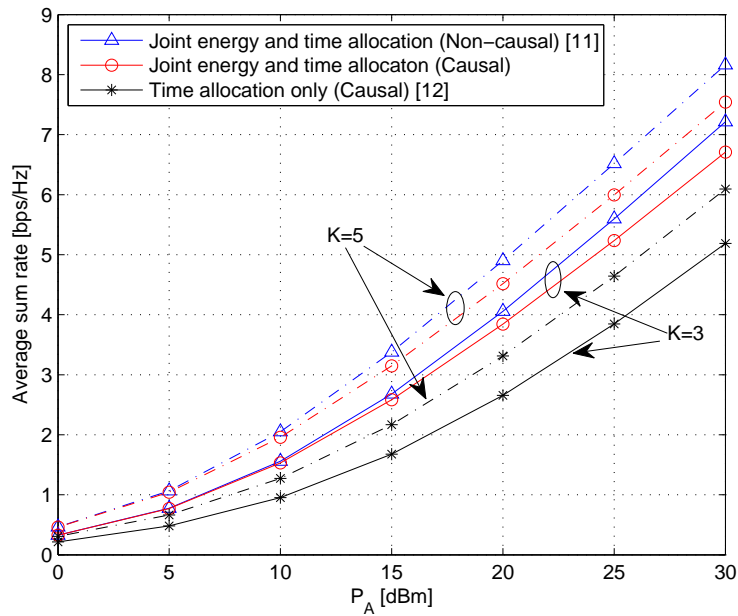}
\end{center}
%\vspace{-7mm}
\caption{Average sum rate performance as a function of $P_{A}$ for infinite capacity energy storage case with $P_{P}=5P_{A}$}
\label{figure:fig1}
\end{figure}

Figure \ref{figure:fig1} depicts the average sum rate performance as a function of $P_{A}$ in the infinite capacity energy storage case. For comparison, we also plot the performance of two conventional schemes in \cite{HJu:14} and \cite{XKang:14}. In \cite{HJu:14}, an ideal non-causal energy system was considered, and the optimal joint energy and time allocation algorithm was proposed for the infinite capacity energy storage case. Meanwhile, \cite{XKang:14} assumed equal power allocation, i.e., $p_{D,i}=P_{A},\forall i$, and provided the optimal time allocation algorithm in the causal energy WPCN systems. In this plot, it is observed that at $P_{A}=30\ \text{dBm}$, the proposed algorithm under the same causal energy scenario provides about $29\%$ and $24\%$ gains over the time allocation scheme in \cite{XKang:14} for $K=3$ and $K=5$, respectively. This implies that the energy allocation optimization offers large performance gains over the system without energy allocation. Note that although the proposed algorithm considers practical causal energy WPCN systems, the performance gap with respect to the non-causal energy system is less than $1\ \text{dB}$ for the average sum rate of $4\ \text{bps/Hz}$.

\begin{figure}
\begin{center}
\includegraphics[width=4.5in]{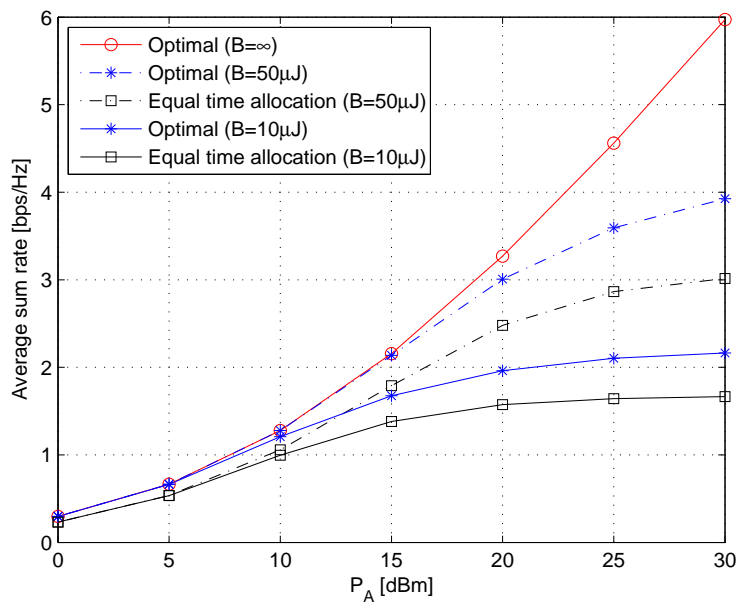}
\end{center}
%\vspace{-7mm}
\caption{Average sum rate performance as a function of $P_{A}$ with $K=3$ and $P_{P}=2P_{A}$}
\label{figure:fig2}
\end{figure}

In Figure \ref{figure:fig2}, we illustrate the average sum rate performance of the WPCN systems as a function of $P_{A}$ with different $B$. Here, the equal time allocation scheme indicates the case with equal time duration for all time slots, i.e., $\tau_{i}=\frac{1}{K+1},\forall i$, and employs the optimal downlink energy allocation policy in (\ref{eq:tilde_eD}). Then, the downlink energy can be computed from (\ref{eq:thm1_1})-(\ref{eq:thm1_3}) with $L=\left\lfloor\frac{P_{A}}{P_{P}(K+1)}\right\rfloor$, where $\lfloor\cdot\rfloor$ stands for the floor operation. We can check in the figure that for a small $P_{A}$, the average sum rate performance in the finite capacity energy storage case is quite similar to that in the infinite capacity energy storage case, since the H-AP cannot transfer enough energy to users regardless of $B$ with a small $P_{A}$. On the other hand, at a high $P_{A}$ regime, the user's energy storage is fully charged for a finite $B$, and thus the user transmits the information signal with the maximum energy $\varepsilon_{U,i}=B_{i}$. Therefore, the average sum rate with a finite $B$ saturates in the high $P_{A}$ regime. Also, it is shown that the performance gap between the proposed optimal algorithm and the equal time allocation scheme increases as $B$ grows, and at $B=50\ \mu\text{J}$ and $P_{A}=30\ \text{dBm}$, the proposed optimal algorithm offers about $30\%$ average sum rate gain.

\begin{figure}
\begin{center}
\includegraphics[width=4.5in]{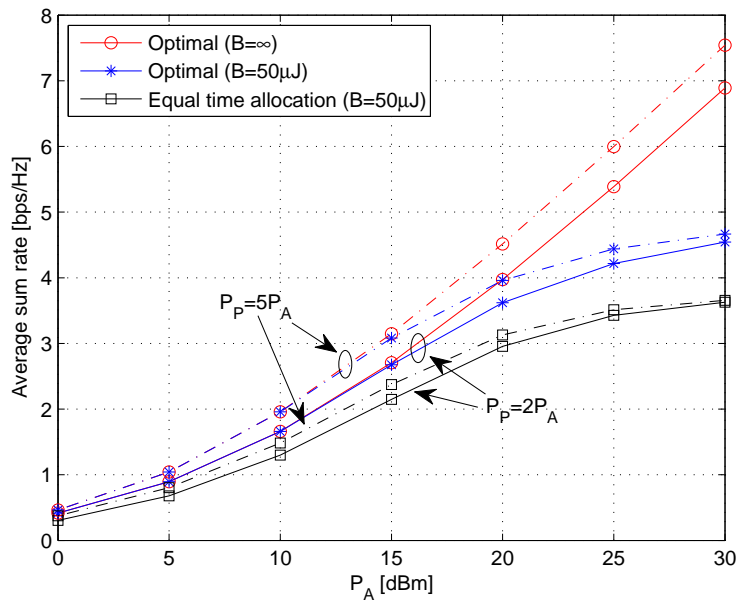}
\end{center}
%\vspace{-7mm}
\caption{Average sum rate performance as a function of $P_{A}$ with $K=5$}
\label{figure:fig2_5}
\end{figure}

Next, with the fixed energy storage capacity $B=50\ \mu\text{J}$, we demonstrate the average sum rate of the proposed optimal algorithm and the equal time allocation method with different peak power constraint $P_{P}$ in Figure \ref{figure:fig2_5}. We can check that as $P_{P}$ gets larger, the average sum rate with $B=\infty$ increases for all $P_{A}$, while the performance is saturated at high $P_{A}$ and $P_{P}$ in the finite capacity energy storage case.

\begin{figure}
\begin{center}
\includegraphics[width=4.5in]{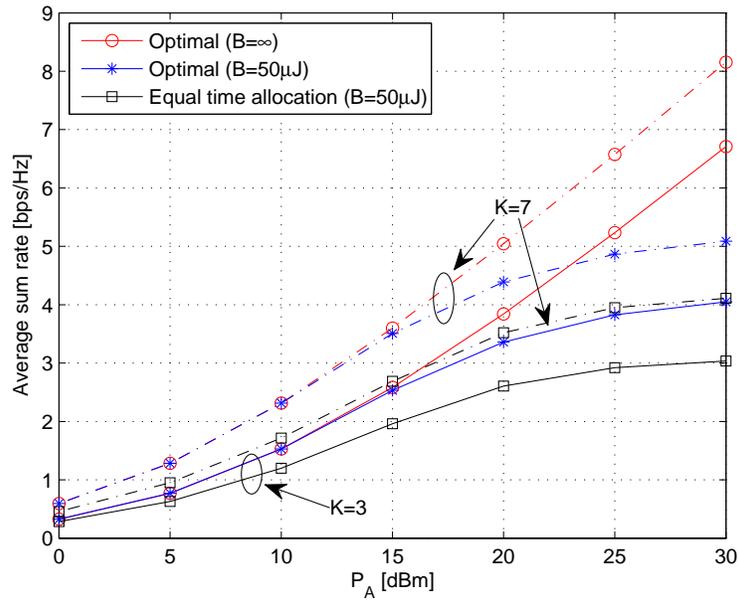}
\end{center}
%\vspace{-7mm}
\caption{Average sum rate performance as a function of $P_{A}$ with $P_{P}=5P_{A}$}
\label{figure:fig3}
\end{figure}

Figure \ref{figure:fig3} exhibits the average sum rate of the WPCN systems with different $K$ as a function of $P_{A}$. We can see that the average sum rates of both the proposed optimal algorithm and the equal time allocation scheme improve with the number of user $K$. With $P_{A}=30\ \text{dBm}$, the proposed optimal algorithm provides $34\%$ and $24\%$ gains over the equal time allocation scheme at $K=3$ and $7$, respectively.

\begin{figure}
\begin{center}
\includegraphics[width=4.5in]{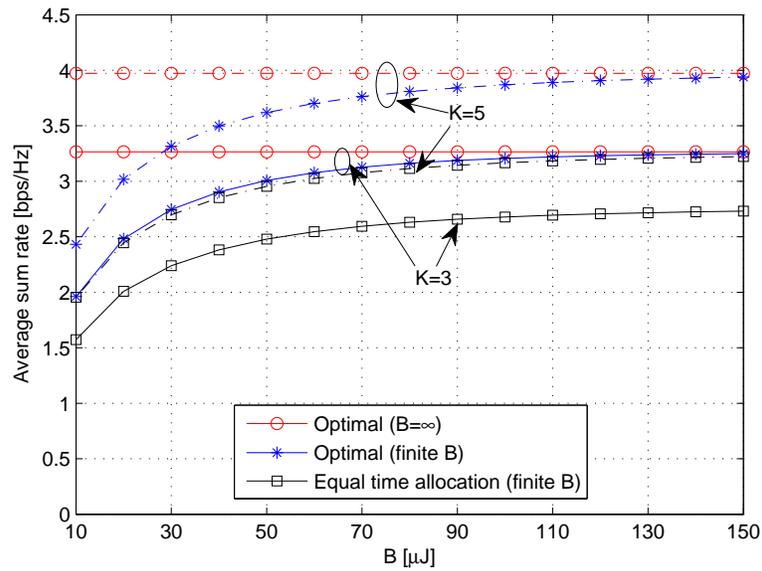}
\end{center}
%\vspace{-7mm}
\caption{Average sum rate performance as a function of $B$ with $P_{A}=20\ \text{dBm}$ and $P_{P}=2P_{A}$}
\label{figure:fig4}
\end{figure}

In Figure \ref{figure:fig4}, the average sum rate performance is presented as a function of $B$ at $P_{A}=20\ \text{dBm}$. With a large $B$, the proposed algorithm for the finite capacity energy storage case provides the performance almost identical to the infinite energy storage case. It is observed that $B=150\ \mu\text{J}$ is enough to achieve the performance upper bound at $P_{A}=20\ \text{dBm}$. On the other hand, the equal time allocation scheme cannot achieve the performance upper bound even if $B$ goes to infinity.

\section{Conclusion}\label{sec:conclusion}
In this paper, we have studied the multi-user WPCN under the causal energy assumption. Joint energy and time allocation problems for maximizing the uplink sum rate have been optimally solved in both the infinite capacity and the finite capacity energy storage cases. First, with the infinite energy storage case, we have derived the optimal downlink energy allocation policy. In this policy, the H-AP consumes all available energy in the first few time slots, and then is turned off during the remaining time slots. Based on this optimal strategy, an analytical solution for the resource allocation problem has been provided. Next, we have proposed the optimal algorithm for the finite capacity energy storage case, which jointly computes energy and time allocation. From the simulation results, we have confirmed that the proposed optimal algorithms provide remarkably enhanced performance compared with conventional techniques.

\begin{appendices}

\section{Proof of Lemma \ref{lemma:lemma1}} \label{appendix:appendixA}
Let us define the optimal solution of problem (\ref{eq:P2}) as $\{\tau_{i}^{\star},\varepsilon_{D,i}^{\star}\}_{i=0}^{K}$, and the corresponding objective value as $R(\{\tau_{i}^{\star},\varepsilon_{D,i}^{\star}\}_{i=0}^{K})$. Suppose that the optimal time allocation solution of problem (\ref{eq:P2}) is given by $\tau_{i}^{\star}=0$ and $\tau_{j}^{\star}>0$ for $j\neq i$. Then, by contradiction, we will show that $\{\tau_{i}^{\star},\varepsilon_{D,i}^{\star}\}_{i=0}^{K}$ is not the optimal solution with $\tau_{i}^{\star}=0$ for $i=0,\cdots,K$. First, we investigate the case of $\tau_{0}^{\star}=0$. Due to the peak power constraint $\varepsilon_{D,0}^{\star}\leq \tau_{0}^{\star}P_{P}=0$, it follows $\varepsilon_{D,0}^{\star}=0$, and thus the uplink rate $R_{1}$ for user $1$ is zero. Now we consider positive numbers $\hat{\tau}_{0}$, $\hat{\tau}_{1}$, $\hat{\varepsilon}_{D,0}$, and $\hat{\varepsilon}_{D,1}$ which fulfill the following conditions:
\bea
    \hat{\tau}_{0}+\hat{\tau}_{1}=\tau_{1}^{\star},\ \hat{\varepsilon}_{D,0}+\hat{\varepsilon}_{D,1}=\varepsilon_{D,1}^{\star},\ \hat{\varepsilon}_{D,0}\leq\hat{\tau}_{0}P_{P},\ \text{and}\ \hat{\varepsilon}_{D,1}\leq\hat{\tau}_{1}P_{P}.\label{eq:appendix1}
\eea

It is worth noting that with $\hat{\tau}_{0}$, $\hat{\tau}_{1}$, $\hat{\varepsilon}_{D,0}$, and $\hat{\varepsilon}_{D,1}$  in (\ref{eq:appendix1}), we can achieve non-zero $R_{1}$ without reducing other users' uplink rate, since the harvested energy of user $j$ for $j=2,\cdots,K$ does not change due to the condition $\hat{\varepsilon}_{D,0}+\hat{\varepsilon}_{D,1}=\varepsilon_{D,1}^{\star}$. Furthermore, we have $\hat{\varepsilon}_{D,0}+\hat{\varepsilon}_{D,1}=\varepsilon_{D,1}^{\star}\leq (\hat{\tau}_{0}+\hat{\tau}_{1})P_{P}=\tau_{1}^{\star}P_{P}$, and thus the positive numbers $\hat{\tau}_{0}$, $\hat{\tau}_{1}$, $\hat{\varepsilon}_{D,0}$, and $\hat{\varepsilon}_{D,1}$ satisfying (\ref{eq:appendix1}) always exist. Therefore, by setting $\hat{\tau}_{j}=\tau_{j}^{\star}$ and $\hat{\varepsilon}_{D,j}=\varepsilon_{D,j}^{\star}$ for $j=2,\cdots,K$, it follows $R(\{\hat{\tau_{j}},\hat{\varepsilon}_{D,j}\}_{j=0}^{K})> R(\{\tau_{j}^{\star},\varepsilon_{D,j}^{\star}\}_{j=0}^{K})$. This contradicts with the assumption that $\{\tau_{j}^{\star},\varepsilon_{D,j}^{\star}\}_{j=0}^{K}$ is optimal.

Second, to show that $\tau_{i}^{\star}=0$ for $i=1,\cdots,K$ do not achieve an optimal solution, we formulate the optimization problem to find a positive solution $\hat{\tau}_{i}$ as
\bea \label{eq:delta}
    \max\limits_{0\leq\tau_{i}\leq\tau_{i+1}^{\star}} \tau_{i}\log\Big(1+\gamma_{i}\frac{\sum_{j=0}^{i-1}\varepsilon_{D,j}^{\star}}{\tau_{i}}\Big)+(\tau_{i+1}^{\star}-\tau_{i})\log\Big(1+\gamma_{i+1}\frac{\sum_{j=0}^{i-1}\varepsilon_{D,j}^{\star}}{\tau_{i+1}^{\star}-\tau_{i}}\Big),
\eea
where we have used the fact $\varepsilon_{D,i}^{\star}=0$. It is worthwhile to note that only $R_{i}$ and $R_{i+1}$ are dependent on $\tau_{i}$, and thus we can improve the sum rate performance by solving (\ref{eq:delta}) without reducing $R_{j},\forall j\neq i,i+1$.

It is known that the optimal $\hat{\tau}_{i}$ must satisfies \cite{HJu:14a}
\bea
    \gamma_{i}\frac{\sum_{j=0}^{i-1}\varepsilon_{D,j}^{\star}}{\hat{\tau_{i}}} = \gamma_{i+1}\frac{\sum_{j=0}^{i-1}\varepsilon_{D,j}^{\star}}{\tau_{i+1}^{\star}-\hat{\tau}_{i}}. \nonumber
\eea
Therefore, we have
\bea \label{eq:delta_opt}
    \hat{\tau}_{i} = \frac{\gamma_{i}}{\gamma_{i}+\gamma_{i+1}}\tau_{i+1}^{\star}.
\eea
Since (\ref{eq:delta_opt}) fulfills $0<\hat{\tau}_{i}<\tau_{i+1}^{\star}$, new solutions $\{\hat{\tau_{i}},\hat{\varepsilon}_{D,i}\}_{i=0}^{K}$ such that $\hat{\tau}_{i+1}=\tau_{i+1}^{\star}-\hat{\tau}_{i}$, $\hat{\tau}_{j}=\tau_{j}^{\star},\forall j\neq i, i+1$ and $\hat{\varepsilon}_{D,j}=\varepsilon_{D,j}^{\star}$ for $j=0,\cdots,K$ increase the sum rate performance, i.e., the assumption $R(\{\hat{\tau_{j}},\hat{\varepsilon}_{D,j}\}_{j=0}^{K})\leq R(\{\tau_{j}^{\star},\varepsilon_{D,j}^{\star}\}_{j=0}^{K})$ is contradiction. This completes the proof.

\section{Proof of Lemma \ref{lemma:lemma2}} \label{appendix:appendixB}
First, we proceed to solve (\ref{eq:dual_decomp}) for $i=0$. In this case, problem (\ref{eq:dual_decomp}) becomes a LP since $\mathcal{J}_{0}$ is an affine function on $\tau_{0}$ and $\varepsilon_{D,0}$. Thus, it is not difficult to show that a solution $\tilde{\varepsilon}_{D,0}$ maximizing $\mathcal{J}_{0}$ is obtained as (\ref{eq:tilde_eD}). Plugging this result into $\mathcal{J}_{0}$, it follows $\mathcal{J}_{0}=(P_{P}\zeta_{0}-\lambda)\tau_{0}$. Since $0\leq\tau_{0}\leq 1$, the optimal $\tilde{\tau}_{0}$ that maximizes $\mathcal{J}_{0}$ is given by $\tilde{\tau}_{0}=1$ if the coefficient of $\tau_{0}$ in $\mathcal{J}_{0}$ is positive, i.e., $P_{P}\zeta_{0}-\lambda>0$. Otherwise, we have $\tilde{\tau}_{0}=0$. Therefore, a solution $\tilde{\tau}_{0}$ can be written by (\ref{eq:tilde_tau}).

Next, we investigate problem (\ref{eq:dual_decomp}) for $i=1,\cdots,K$. Similar to the case of $i=0$, a solution $\tilde{\varepsilon}_{D,i}$ for $i=1,\cdots,K$ is determined by (\ref{eq:tilde_eD}), since $\mathcal{J}_{i}$ is an affine function of $\tilde{\varepsilon}_{D,i}$. Substituting (\ref{eq:tilde_eD}) into $\mathcal{J}_{i}$ yields
\bea
    \mathcal{J}_{i}=\tau_{i}\log\bigg(1+\frac{g_{U,i}}{\sigma_{i}^{2}}\frac{\varepsilon_{U,i}}{\tau_{i}}\bigg)+(P_{P}\zeta_{i}-\lambda)\tau_{i}-\beta_{i}\varepsilon_{U,i}. \label{eq:Ji}
\eea
Then, by using the zero gradient condition $\frac{\partial\mathcal{J}_{i}}{\partial\tau_{i}}=0$ and $\frac{\partial\mathcal{J}_{i}}{\partial\varepsilon_{U,i}}=0$, we have
\bea
    \log\Big(1+\frac{g_{U,i}}{\sigma_{i}^{2}}\frac{\varepsilon_{U,i}}{\tau_{i}}\Big) + \frac{1}{1+\frac{g_{U,i}}{\sigma_{i}^{2}}\frac{\varepsilon_{U,i}}{\tau_{i}}}\!\!\!&=&\!\!\!1+\lambda-P_{P}\zeta_{i}, \label{eq:zero_grad22} \\
    \frac{\frac{g_{U,i}}{\sigma_{i}^{2}}}{1+\frac{g_{U,i}}{\sigma_{i}^{2}}\frac{\varepsilon_{U,i}}{\tau_{i}}}\!\!\!&=&\!\!\!\beta_{i}. \label{eq:zero_grad11}
\eea
It can be shown that a solution of the equation (\ref{eq:zero_grad22}) does not exist if $P_{P}\zeta_{i}-\lambda >0$ since $\frac{g_{U,i}}{\sigma_{i}^{2}}\frac{\varepsilon_{U,i}}{\tau_{i}}\geq0$ in general. Therefore, from (\ref{eq:zero_grad22}), we can obtain a solution $\tilde{\tau}_{i}$ for $i=1,\cdots,K$ as in (\ref{eq:tilde_tau}), and the dual variables must satisfy the condition $P_{P}\zeta_{i}-\lambda \leq0$.

Also, combining (\ref{eq:zero_grad11}) and the constraint $0\leq\varepsilon_{U,i}\leq B_{i}$, a solution $\tilde{\varepsilon}_{U,i}$ can be expressed by
\bea
    \tilde{\varepsilon}_{U,i}\!\!&=&\!\!\min\bigg\{\bigg(\frac{1}{\beta_{i}}-\frac{\sigma_{i}^{2}}{g_{U,i}}\bigg)\tilde{\tau}_{i},B_{i}\bigg\}. \label{eq:tilde_eU_app}
\eea
Here, to ensure $\tilde{\varepsilon}_{U,i}\geq 0$, the dual variable $\beta_{i}$ must be upper bounded by $\beta_{i}\leq g_{U,i}/\sigma_{i}^{2}$. Then, from (\ref{eq:tilde_tau}) and (\ref{eq:tilde_eU_app}), we can see that $\tilde{\varepsilon}_{U,i}$ is equal to a solution of the following fixed point equation:
\bea
    \varepsilon_{U,i}=\min\bigg\{\frac{-g_{U,i}/\sigma_{i}^{2}}{1/b_{i}+1}\bigg(\frac{1}{\beta_{i}}-\frac{\sigma_{i}^{2}}{g_{U,i}}\bigg)\varepsilon_{U,i},B_{i}\bigg\}. \label{eq:fixed_point}
\eea
With any feasible initial point $0 \leq z_{i} \leq B_{i}$, a solution of (\ref{eq:fixed_point}) is computed as (\ref{eq:tilde_eU}). This completes the proof.
\end{appendices}

\nocite{*}
\bibliography{HLee_WPCN_onecol}
\bibliographystyle{ieeetr}

\end{document}